\documentclass[lettersize,journal]{IEEEtran}
\usepackage{amsmath,amsfonts}
\usepackage{algorithm}
\usepackage{array}
\usepackage[caption=false,font=normalsize,labelfont=sf,textfont=sf]{subfig}
\usepackage{textcomp}
\usepackage{stfloats}
\usepackage{url}
\usepackage{verbatim}
\usepackage{graphicx}
\usepackage{cite}
\usepackage{color}

\usepackage{pifont}

\usepackage{algorithm}
\usepackage{algpseudocode}
\usepackage{amsmath}
\usepackage{amsfonts}
\usepackage{amssymb}
\usepackage{graphics}
\usepackage{epsfig}

\usepackage{url}
\usepackage{graphicx}
\usepackage{multicol}  
\usepackage{multirow}  
\usepackage{subfig}  
\usepackage{booktabs}
\usepackage{pgfplots}
\usepackage{ragged2e}

\usepackage{colortbl}  %彩色表格需要加载的宏包
\usepackage{xcolor}

\usepackage{enumitem}

\usepackage[utf8]{inputenc} 
\begin{document}

\title{Large Language Model Enhanced Text-to-SQL Generation: A Survey}

\author{Xiaohu Zhu, Qian Li, Lizhen Cui, Yongkang Liu}

% The paper headers
% \markboth{Journal of \LaTeX\ Class Files,~Vol.~14, No.~8, August~2021}%
% {Shell \MakeLowercase{\textit{et al.}}: A Sample Article Using IEEEtran.cls for IEEE Journals}

% \IEEEpubid{0000--0000/00\$00.00~\copyright~2021 IEEE}
% Remember, if you use this you must call \IEEEpubidadjcol in the second
% column for its text to clear the IEEEpubid mark.

\maketitle

\begin{abstract}
Text-to-SQL translates natural language queries into Structured Query Language (SQL) commands, enabling users to interact with databases using natural language. Essentially, the text-to-SQL task is a text generation task and its development primarily dependent on changes in language models. Especially with the rapid development of Large Language Models (LLMs), the pattern of text-to-SQL has undergone significant changes.
Existing survey work mainly focuses on rule-based and neural-based approaches, still lacking a survey of Text-to-SQL with LLMs. 
In this paper, we survey the large language model enhanced text-to-SQL generations, classifying them into prompt engineering, fine-tuning, pre-trained and Agent groups according to training strategies.
And we also summarize datasets and evaluation metrics comprehensively.
This survey could help people better understand the pattern, research status, and challenges of LLM-based text-to-SQL generations.

% used in the field. In addition, we highlight the emerging trend of deploying LLM in databases, emphasizing the potential of LLM to improve query accuracy and accessibility in complex real-world scenarios. 
% Traditional text-to-SQL systems, including rule-based engineering approaches and deep neural networks, have made significant progress. In recent years, Pre-training models have been introduced and applied to this task, bringing significant performance improvements. However, with the increasing complexity of modern databases and user queries, Pre-training models still have limitations in handling complex SQL generation with parameter constraints. 
% This survey provides a comprehensive overview of the evolution of text-to-SQL systems, tracing the development from traditional approaches to the latest advances driven by Large Language Models (LLMs). This survey aims to provide insights into the current state and future direction of text-to-SQL research and to contribute to the advancement of intelligent data querying techniques.

\end{abstract}

\begin{IEEEkeywords}
Text-to-SQL, Large Language Models,
Prompt Engineering,
Fine-Tuning,
Database Querying
% Semantic Parsing，
% In-Context Learning，
% Evaluation Metrics，
% Dataset Benchmarks，
% SQL Generation，
% Query Optimization，
% Retrieval-Augmented Generation (RAG)
\end{IEEEkeywords}

\section{Introduction}
\IEEEPARstart{D}{ata} has become a crucial production factor \cite{xu2021datafactor,Zhong2023} in the productive life of human activities. With the proliferation of electronic devices, there have been more and more databases appearing, storing massive information from all sorts of areas\cite{TheEconomicsandImplicationsofDataAnIntegratedPerspective, su15097287}. However, the threshold for learning database query language, such as SQL, is relatively high for ordinary people. Even for practitioners, it is more troublesome to write a large number of query statements with guaranteed correctness for different domain databases and application scenarios.  
To lower the barriers of using database queries, text-to-SQL task translates natural language queries into Structured Query Language (SQL) commands, enabling users to interact with databases using natural language. 

\begin{figure}[t]
    \centering
    \includegraphics[width=0.8\columnwidth]{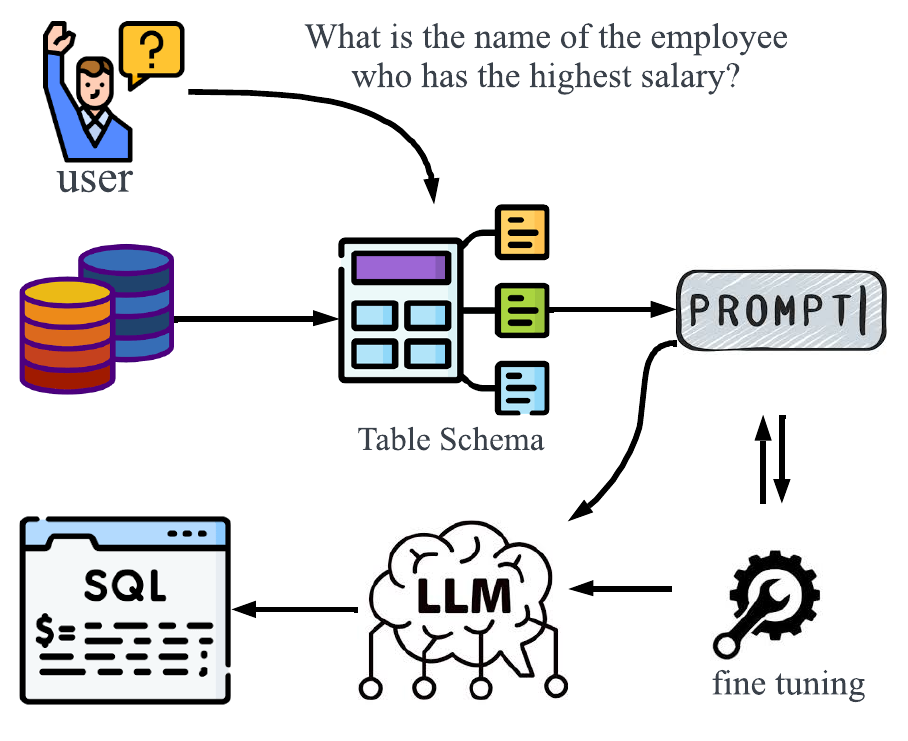} 
    \caption{Flowchart of the Text-to-SQL. The flowchart illustrates the process where user questions and the database schema are first collected. These inputs are then processed through prompt engineering and fine-tuning techniques before being passed to a large language model (LLM). The LLM generates the corresponding SQL query based on the refined inputs, allowing for accurate query formulation based on natural language input.} 
    \label{fig:flowchart}  
\end{figure}

Fig.\ref{fig:flowchart} gives an example of a text-to-SQL task. Given a natural language question \textbf{Q} and a database schema \textbf{S}:

\noindent \textbf{Q:}  \textit{What is the name of the employee with the highest salary?}

\noindent \textbf{S: } \textit{Table: Employees (ID, Name, Salary)}

\noindent The goal of text-to-SQL is to generate a SQL query $\hat{Y}$,

 \ \ \ \ \ \ \ \ \ \ \textit{SELECT Name FROM Employees}

 \ \ \ \ \ \ \ \ \ \ \textit{ORDER BY Salary DESC LIMIT 1;}

\noindent After converting text into SQL language, we can search for relevant knowledge from databases, thus breaking down the barriers between natural language and structured data \cite{liu2020fareffectivecontextmodeling}.

% In this analysis, we will explore the process of the model deciding on a precise question statement by examining the inquiry and database schema. It then performs a query for the database and subsequently gets the answer (see Fig. \ref{fig:flowchart}).

% Thus, we highlight the relevance of AI and NLP and describe the technology that is quietly changing the data interaction script - transforming nature languages into effective database queries. SQL is the leading standard for managing it. Additionally, this transition opens up enormous potential for the development of data-driven applications. Here, the SQL programming language is hierarchically developed based on a natural language text, a sub-task in the topic of Semantic Parsing, as elaborated in the work of Liu et al.\cite{liu2020fareffectivecontextmodeling}.

% Essentially, the text-to-SQL task is a text generation task, so its development primarily dependent on changes in language models.

The history of Text-to-SQL goes back to 1973 when \cite{woods1972lunar} developed a system called the LUNAR system, which was primarily used to answer questions related to rocks brought back from the Moon. The earliest researches are mostly based on fine-designed rules \cite{li2014constructing}, which are suitable for uncomplicated or specific scenarios. As the amount and domains of data grow exponentially, it becomes expensive for these rule-based methods. Deep neural networks began to take center stage as foundational approaches, such as LSTM-based\cite{10.1162/neco.1997.9.8.1735} and Transformer-based\cite{NIPS2017_3f5ee243} methods. However, they are confronted with problems such as data sparsity and generalization issues. 

% Previous research and investigation have predominantly concentrated on the extraction of natural language (NL) questions to structured query language (SQL) patterns and their generalization through the training of encoder-decoder models using a Text-to-SQL corpus\cite{qin2022surveytexttosqlparsingconcepts, shi2024surveyemployinglargelanguage, deng-etal-2022-recent}, such as LSTM\cite{10.1162/neco.1997.9.8.1735} and Transformer\cite{NIPS2017_3f5ee243}. The utilization of a sizable corpus for training facilitates the automatic acquisition of the mapping relationship between natural language and SQL, obviating the need for manually crafted rules. Pre-training models have become the state-of-the-art methods at this moment\cite{bei2023cpdgcontrastivepretrainingmethod}.

% However, traditional models like LSTM-based and early Transformer-based models face limitations in understanding complex schema and generating highly accurate SQL queries. To address this, researchers introduced fine-tuning techniques to improve model adaptability to specific datasets and query structures. By fine-tuning, models such as BERT and T5 achieved greater success in capturing the relationships between natural language and SQL in diverse and complex domains. Prompt Engineering techniques further enhanced these models' performance by carefully crafting input prompts, guiding the models to generate more accurate SQL outputs.

Recently, with the significant improvement of inference and generalization abilities in Large Language Models (LLMs), many works use LLMs to generate SQL queries correctly, and have achieved greater abilities to understand natural language than previous approaches \cite{hong2024nextgenerationdatabaseinterfacessurvey}. For instance, ChatGPT-4\cite{OpenAI_GPT4_2023} has achieved the top performance on the Spider\cite{yu-etal-2018-spider} dataset, setting the new standard for execution accuracy. Existing survey work mainly focuses on rule-based and neural-based approaches, still lacking a survey of Text-to-SQL with LLMs.

% In comparison to traditional Transformer-based models and pre-trained models that rely on fine-tuning, LLMs offer a more flexible approach to Text-to-SQL, often requiring fewer task-specific modifications due to their extensive pre-training on diverse datasets. LLMs also open up new opportunities for handling open-domain questions and schema variations, further improving the field’s performance standards.

% In parallel, the task of the Text-to-SQL domain has been better developed with the introduction of Pre-trained Language Models (PLMs). PLMs enable the pre-training of language models on large-scale unlabeled data, allowing them to capture semantic and other utterance information in a comprehensive and rich way. This gives the models better generalization capabilities, and the use of pre-trained models allows researchers to automatically generate mappings from natural language to SQL without having to manually design specific rules, drastically reducing the problems of data scarcity and annotation costs.

% Beyond PLMs, another emerging direction involves the integration of advanced prompt-based techniques. By designing effective prompts, researchers have leveraged both pre-trained models and LLMs to optimize SQL generation tasks, particularly in zero-shot and few-shot learning scenarios, where models are required to generate accurate SQL queries without extensive task-specific training.

\begin{figure*}[htbp]
    \centering
    \includegraphics[width=\textwidth]{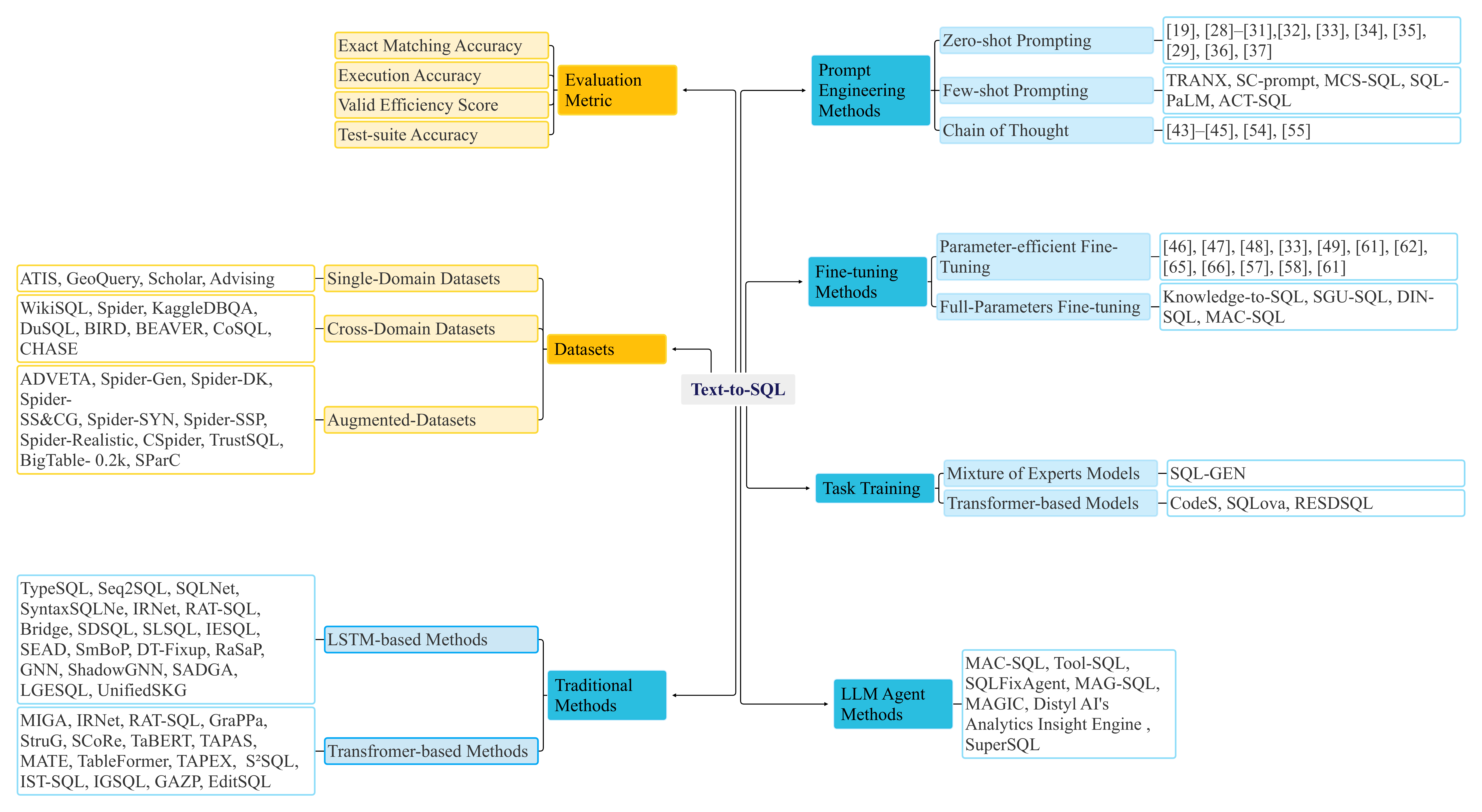} 
    \caption{The overview of the text-to-SQL metrics, datasets, and methods.} 
    \label{fig:overview}  
\end{figure*}

% The purpose of this paper is to give a comprehensive overview of the development of natural language to database queries. Much of the research in this area has made far-reaching developments, from the initial simple rule-based approaches to the previously SOTA Deep learning-based systems. These systems initially relied on traditional LSTM-based and Transformer-based models, which introduced significant improvements in mapping natural language to SQL. However, with the rise of pre-trained language models (PLMs) and, more recently, large language models (LLMs), the field has experienced a paradigm shift in how SQL queries are generated. In this paper, in addition to presenting the development process of Text-to-SQL, we will present the latest research in the field, integrating the latest LLM-based approaches, including:

In this paper, we survey the large language model enhanced text-to-SQL generation methods, classifying them into the prompt, fine-tuning, task-training, and agent according to training strategies, as shown in Fig.\ref{fig:pdfpage},
\begin{itemize}
    \item \textbf{Prompt:} (No training) They use well-designed prompts to guide LLMs to generate more accurate SQL queries, enabling powerful LLMs to generate SQL queries from zero-shot \cite{Rajkumar2022EvaluatingTT,liu2023comprehensiveevaluationchatgptszeroshot} or few-shot \cite{brown2020languagemodelsfewshotlearners,Lee2024MCSSQL} examples. 

    \item \textbf{Fine-Tuning:} (Training from pretrained LLMs) They finetune the LLM models to adapt to the text-to-SQL task, including full-parameters fine-tuning \cite{Hong2024KnowledgeToSQL,Zhang2024Structure} and parameter-efficient fine-tuning \cite{song2022learningnoisylabelsdeep,Gao2023TexttoSQLEmpowered}. 

    \item \textbf{Task-Training:} (Training from scratch) They train a task-specific text-to-SQL model with training strategies similar to LLMs, such as Transformers \cite{Li2024CodeS,Fu2023MIGA} and mixture of experts \cite{pourreza2024sqlgenbridgingdialectgap}.

    \item \textbf{LLM Agent:} (Training with multiple agents and external tools) They collaborate with multiple intelligences, dynamically generate and correct SQL queries, handle database matching issues, and improve query accuracy and execution through external tools \cite{wang2024macsqlmultiagentcollaborativeframework,wang2024tool}. 
    
\end{itemize}

% \textbf{Methods}: 
% We provide a comprehensive review of the various methods in the Text-to-SQL domain, from the traditional approaches to the latest LLM-based technological developments. In the early stages, methods like Seq2Seq models, LSTM-based approaches, and attention-based Transformers laid the foundation for Text-to-SQL. As the field progressed, LSTM-based and Transformer-based methods changed the landscape, allowing for better generalization and reduced reliance on large labeled datasets. 
 
Meanwhile, we summarize the datasets and metrics for the large language model enhanced text-to-SQL generations.
For datasets, we explore the characteristics of datasets in terms of data source scenarios, number of tables, SQL complexity, and number of conversation rounds by systematically combing single-domain, cross-domain, and augmented datasets, and analyze the challenges and limitations of these datasets in real-world applications.
For metrics, we consider exact matching accuracy, execution accuracy, valid efficiency and test-suite accuracy in handling single-turn, multi-turn interactions. 

In the following, we first give the preliminaries in Section II, then introduce the metrics and datasets in Section III, third propose detailed descriptions of the method in Section IV, and give a conclusion and future work at the final.

% Developments in Text-to-SQL technology will continue to focus on improving the accuracy, generalization capabilities, and efficiency of the models. The future of Text-to-SQL will likely be shaped by more advanced LLMs, with improved context handling and the ability to dynamically adjust to user intent during multi-round conversations. Moreover, we expect to see more innovative approaches in Prompt Engineering that will guide models to generate SQL queries more effectively, even in the absence of large domain-specific datasets. This may include further optimizing the performance of large-scale language models when generating complex SQL queries, and developing smarter Prompt Engineering approaches. Furthermore, the integration of LLM Agent systems, which offer task management and multi-round interaction.

\section{Preliminaries}
The Text-to-SQL systems enable users to input questions directly in natural language, and the system will automatically generate corresponding SQL query statements.
\subsection{The Text-to-SQL problem}
Given a natural language question \(Q\) and a database \(S\), The objective of Text-to-SQL tasks is to generate an accurate SQL query \(\hat{Y}\) that retrieves the desired output from databases \(D\). This can be conceptualized as a sequence-to-sequence\cite{see2017pointsummarizationpointergeneratornetworks} problem:

\begin{itemize}
    \item \textbf{Input:}
    \begin{itemize}
        \item a Natural Language problem:\(Q = (q_1, q_2, \ldots, q_n)\), where \(q_i\) represents the \(i\)-th token in the question.
        \item A database schema:\(S = \{T_1, T_2, \ldots, T_m\}\), where \(T_i\) represents the \(i\)-th table in the database, and each table \(T_i\) includes columns \(C_{i1}, C_{i2}, \ldots, C_{ip}\).
    \end{itemize}
    \item \textbf{Output:}
    \begin{itemize}
        \item A SQL query:\(\hat{Y} = (\hat{y}_1, \hat{y}_2, \ldots, \hat{y}_k)\), where \(\hat{y}_i\) represents the \(i\)-th token in the generated query.
    \end{itemize}
\end{itemize}

The task can be expressed as finding the most probable SQL query \(\hat{Y}\) given a natural language question \(Q\) and a database schema \(S\):

\[
\hat{Y} = \arg\max_Y P(Y \mid Q, S)
\]

\subsection{Mythology}
To solve this problem, modern methodologies typically employ deep learning models\cite{10.1007/s00778-022-00776-8}, particularly Encoder-Decoder (ED) architectures. Presented below is a high-level overview of the process:
\begin{itemize}
    \item \textbf{Encoding:}
    The encoder processes the input question \(Q\) and the schema \(S\) to create a contextual representation. This can be represented as:

    \[
    h = \text{Encoder}(Q, S)
    \]

    where \(h\) is the hidden state or contextual representation generated by the encoder.
    
    \item \textbf{Decoding:}
    The decoder generates the SQL query \(\hat{Y}\) token by token based on the encoded representation \(h\). The probability of each token in the SQL query can be calculated as:

    \[
    P(Y \mid Q, S) = \prod_{i=1}^{k} P(\hat{y}_i \mid h, \hat{y}_{1:i-1})
    \]

    where \(\hat{y}_{1:i-1}\) are the previously generated tokens, \(P(\hat{y}_i \mid h, \hat{y}_{1:i-1})\) represents the probability of the\(i\)th token given the context and previous tokens.

    \item \textbf{Optimization:}
    The model is trained to maximize the likelihood of the correct SQL query \(Y\)  given the training data. The loss function typically used is the negative log-likelihood of the correct query:

    \[
    \mathcal{L} = - \sum_{i=1}^{k} \log P(y_i \mid h, y_{1:i-1})
    \]

    where \(y_i\) are the tokens of the ground truth SQL query.
\end{itemize}

\subsection{Challenges}
%%%%%%%%%%%%%%%%%
% \subsubsection{Ambiguity and Variability:}
Ambiguity presents one of the most prevalent and intractable problems in Natural Language Processing, it denotes the phenomenon wherein a single and the same linguistic form may be interpreted in more than one way\cite{hong2024betterquestiongenerationqabased,zheng2023editfactualknowledgeincontext}.

\subsubsection{\textbf{Word Segmentation and Sense Ambiguity}} 
Word Segmentation Ambiguity refers to the phenomenon of different meanings when words are combined from characters. For Indo-European languages, most words are separated by spaces or punctuation. However, in languages such as Chinese and Japanese, words are usually not separated by a space or a punctuation mark. Consequently, ambiguity arises when these consecutive characters are attempting to segment into words. 

Word sense ambiguity denotes the phenomenon where a word shares identical orthography while possessing distinct semantic interpretations in linguistic terms. 
For example, Bank noun:
\textit{(1) sloping land (especially the slope beside a body of water
“they pulled the canoe up on the bank”
(2) a financial institution that accepts deposits and channels the money into lending activities
“he cashed a check at the bank”}

\subsubsection{\textbf{Database size and diversity}} 
Realistic databases often contain hundreds of tables and columns, and the relationships between different tables can be very complex. Due to the sheer size of the database schema, Text-to-SQL systems are often unable to incorporate all relevant table structure information in a single prompt. This poses a challenge to the model because without the complete schema context, it is difficult for the model to generate correct SQL queries. To further complicate matters, databases in different domains may have completely different naming conventions, formats, and table structures. For example, column names in some databases may not have intuitive meanings (e.g., “col1”, “data1”), or even have a large number of abbreviations or naming ambiguities, which requires the model to have good reasoning capabilities to correctly understand the relationships between tables and columns. In addition, the data types and formats in the database are diversified, for example, date data may have different representations (e.g., “2024-01-01” or “year2024”), which further increases the difficulty of data parsing.
\subsubsection{\textbf{Complexity of SQL queries}
} 
The complexity of SQL queries is usually related to the query structure, involving operations such as joins of multiple tables, nested subqueries, and complex conditional filtering. For example, a query may contain both SELECT, JOIN, GROUP BY, HAVING, and nested WHERE conditions, which requires the model to be able to understand and generate complex SQL statements efficiently. Particularly in multi-table queries, the model must be able to infer table-to-table relationships (e.g., foreign keys) and ensure that the generated queries are logically correct. For example, SQL with nested subqueries requires the model to have the ability to handle hierarchical relationships, while conditional filtering and aggregation operations require the model to be able to generate precise expressions based on different column types and values. In addition, certain queries in a given domain may require the use of specific SQL functions or operations (e.g., regular expression matching), which further complicates the SQL generation process.

In a paragraph or chapter consisting of multiple sentences, various kinds of ambiguities remain, such as \textit{referential ambiguities} and \textit{ellipsis ambiguities}. Referential ambiguity is the possibility of ambiguity in the events referred to by pronouns (e.g., I, you, he) and pronoun phrases.

\subsubsection{\textbf{Pragmatic Ambiguity}} 

Pragmatic Ambiguity pertains to ambiguity due to context, speaker attributes, scene, and other situational pragmatic aspects. A sentence may elicit varying interpretations across different contexts. For example, the following example shows that the same sentence can yield different meanings based on different scenes.

Sentence: \textit{Do you know how to get to Fifth Avenue?}

\textit{If the speaker is a tourist and the person speaking is a policeman, the meaning of the sentence is to ask for directions.}
\textit{If the speaker is also a tourist, but the person speaking is a taxi driver instead, then the meaning of the sentence is to ask for a ride to Fifth Avenue.}
\subsubsection{\textbf{Robustness and Efficiency}} 
Users may input queries with spelling mistakes, grammatical errors, or incomplete sentences.  In real-world applications, users often provide imperfect or ambiguous queries due to human error or lack of domain knowledge. The model should be robust enough to deal with such inputs.

Correctly mapping valid information in natural language to corresponding parts of the database is one of the most critical steps. The model needs to understand information such as the structure and constraints of the schema. In addition to this, the structure of the schema format will change in different databases, and the model must have good generality.

One of the basic requirements of the model is the correctness of the query generated by SQL. Due to the special characteristics of SQL, even a small error can cause the entire query to fail to produce the correct answer.

Also, the execution efficiency of SQL queries is a key consideration in practical applications. Even if the model is able to generate accurate SQL queries, if the execution efficiency of these queries is too low, especially on large-scale databases, this will affect the practicality of the system. A good Text-to-SQL system should not only generate correct queries, but also ensure that these queries can be executed quickly and efficiently in the actual database to reduce system response time and improve user experience.

\section{Metrics and Datasets}
\subsection{Evaluation Metric}
To evaluate the performance of Text-to-SQL models, several key metrics are employed to assess the accuracy, execution effectiveness, and overall efficiency of the generated SQL queries. These metrics provide a comprehensive understanding of how well a model translates natural language questions into valid SQL statements, ensuring both syntactic correctness and proper execution on the given database schema. In this section, we introduce four commonly used evaluation metrics: Exact Matching Accuracy (EM), Execution Accuracy (EX), Valid Efficiency Score (VES), and Test-suite Accuracy (TS). Each of these metrics plays a crucial role in highlighting different aspects of model performance, from syntactic correctness to real-world execution efficiency.

\subsubsection{Exact Matching Accuracy(EM)}
Exact Matching Accuracy \cite{yu-etal-2018-spider}requires that the SQL statement generated by the model must be exactly the same as the ground-truth answers. It is a critical metric for evaluating the performance of Text-to-SQL. This stringent metric requires an identical query due to the diversity of syntax in SQL statements. Completing the same results may not always be possible to determine the correct SQL query for the same task uniquely.

\[
\text{Exact Matching Accuracy} = \frac{1}{N} \sum_{i=1}^{N} \mathbb{I} \left( \hat{Y}_i = Y_i \right)
\]

where:

\begin{itemize}
  \item \textbf{Total number of queries} \( N \): This represents the total number of natural language questions used in the evaluation.
  \item \textbf{Generated SQL query} \( \hat{Y}_i \): This is the SQL query generated by the model for the $i$-th natural language question.
  \item \textbf{Reference SQL query} \( Y_i \): This is the correct SQL query for the $i$-th natural language question, serving as the reference standard.
  \item \textbf{Indicator function} \( \mathbb{I}(\cdot) \): If the generated SQL query \( \hat{Y}_i \) exactly matches the reference SQL query \( Y_i \), the indicator function \( \mathbb{I}(\cdot) \) equals 1; otherwise, it equals 0.
\end{itemize}

\subsubsection{Execution Accuracy(EX)}

\[
\text{Execution Accuracy} = \frac{1}{N} \sum_{i=1}^{N} \mathbb{I} \left( f(Q_i, S_i) = A_i \right)
\]

\begin{itemize}
  \item \( N \) is the total number of queries.
  \item \( Q_i \) denotes the $i$-th natural language question.
  \item \( S_i \) denotes the database schema corresponding to the $i$-th question.
  \item \( A_i \) denotes the reference answer for the $i$-th question.
  \item \( f(Q_i, S_i) \) denotes the result returned by executing the SQL query generated by the model for the $i$-th question on the database schema \( S_i \).
  \item \( \mathbb{I}(\cdot) \) is the indicator function, which equals 1 if the condition inside is true, and 0 otherwise.
\end{itemize}

    \subsubsection{Valid Efficiency Score}
Valid Efficiency Score (VES) is a common evaluation metric used in the text-to-SQL field to assess the performance of models. It considers both the correctness and the efficiency of the generated SQL queries. The metric accounts for the validity of the SQL query (whether it executes correctly and produces the correct result) and its execution efficiency (how fast it runs).

The formula typically consists of two components:

Query Validity (Correctness): Evaluates whether the generated SQL can be successfully executed and returns the same result as the ground truth SQL query.
Query Efficiency: Measures how efficiently the generated SQL query is executed compared to the ground truth.
\[
VES = \frac{1}{N} \sum_{i=1}^{N} \left( \mathbb{I}(Q_i^{\text{gen}} = Q_i^{\text{gold}}) \cdot \frac{T_{\text{gold}}}{T_{\text{gen}}} \right)
\]

where:
\begin{itemize}
  \item \( N \) represents the total number of queries;
  \item \( Q_i^{\text{gen}} \) denotes the generated SQL query for the \(i\)-th example;
  \item \( Q_i^{\text{gold}} \) denotes the ground truth SQL query for the \(i\)-th example;
  \item \( \mathbb{I}(\cdot) \) is the indicator function, equal to 1 if the generated SQL is equal to the ground truth, and 0 otherwise;
  \item \( T_{\text{gold}} \) is the execution time of the ground truth SQL query;
  \item \( T_{\text{gen}} \) is the execution time of the generated SQL query.
\end{itemize}
\subsubsection{Test-suite Accuracy (TS)}
Test-suite Accuracy (TS) is a key evaluation metric that is designed to test the performance of models on a diverse and concentrated set of test databases. This metric constructs a small, focused database test suite from a large collection of randomly generated databases, ensuring that the suite has a high code coverage rate for accurate SQL queries. By testing the model's performance on this suite, Test-suite Accuracy measures how well the model predicts SQL queries that produce the correct results across various database scenarios. The goal is to measure the strict upper limit of semantic accuracy, as it evaluates the performance not only in terms of SQL structure but also execution outcomes.

\[
\text{Test-suite Accuracy} = \frac{1}{N} \sum_{i=1}^{N} \mathbb{I} \left( f(Q_i, D_i) = R_i \right)
\]

where:
\begin{itemize}
  \item \( N \) represents the total number of queries in the test suite.
  \item \( Q_i \) is the SQL query generated by the model for the \( i \)-th example.
  \item \( D_i \) represents the corresponding database for the \( i \)-th query.
  \item \( R_i \) is the expected result when executing the reference SQL query on \( D_i \).
  \item \( f(Q_i, D_i) \) is the result produced by executing the model-generated SQL query on database \( D_i \).
  \item \( \mathbb{I}(\cdot) \) is the indicator function, which equals 1 if the generated query result matches the expected result \( R_i \), and 0 otherwise.
  
\end{itemize}

\subsection{Datasets}
Datasets are a fundamental component of training Text-to-SQL tasks. Training the system with a large corpus allows it to automatically acquire the mapping relationship between natural language and SQL without relying on hand-written rules. The dataset for Text-to-SQL is usually manually labeled with natural language questions and corresponding SQL queries. A natural language question is a question restricted to the domain in which the database data is located and whose answer comes from its database. In essence, the question describes a SQL query. Executing the SQL query yields the answer to the question from its database\cite{li-etal-2024-multisql}. Table \ref{dataset} provides an overview of commonly used Text-to-SQL datasets, summarizing key features such as dataset size, interaction type, and domain coverage, which are essential for evaluating the generalization capability of Text-to-SQL models.
Datasets in this domain typically have the following characteristics.

Single/Cross Domain: database data source scenarios, according to the number of scenarios involved, can be divided into single fields and multiple fields\cite{app13042262}, such as catering data and tourist attractions for two fields.

Number of dialogue rounds: according to the number of dialogue rounds required for complete SQL generation, the dataset is divided into single and multiple rounds.

SQL Complexity: Based on the SQL complexity corresponding to the natural language problem, the dataset is classified into simple and complex problems, where the problem complexity is determined by the number of keywords, nesting level, and number of clauses.

\begin{table*}[t]
	\centering
        \caption{   
        Dataset statistics. Provides a summary of key Text-to-SQL datasets, highlighting attributes such as size, interaction type, domain coverage, language, release year, and link 
	}
	\resizebox{1\textwidth}{!}{
	%	\resizebox{19cm}{1.6cm}{x
	\begin{tabular} {l  ccc cccc >{\centering\arraybackslash}p{8cm}}
	\toprule
        
        \multirow{2}{*}[-5pt]{Name} &
	\multicolumn{3}{c}{Number} &
        \multirow{2}{*}[-5pt]{Turn} &
        \multirow{2}{*}[-5pt]{Type} &
        \multirow{2}{*}[-5pt]{Language} &
        \multirow{2}{*}[-5pt]{Year} &
        \multirow{2}{*}[-5pt]{Link} \\
        \cmidrule(lr){2-4}
        
        & Train & Valid & Test & & & & & \\
        % Name & Example & Number(train, valid, test) & Type & Language & Year & Link & \\ 
        \midrule  
         ATIS  \cite{price-1990-evaluation}
     
        & 4473 & 497 & 448 & Single-Turn & Single-Domain & English & 1990 & \url{https://github.com/howl-anderson/ATIS_dataset}\\
        
        \midrule

        GeoQuery \cite{10.5555/1864519.1864543}
         
           & 600 &  -& 280 & Single-Turn & Single-Domain & English & 1996 & https://www.cs.utexas.edu/~ml/nldata/geoquery.html \\
        \midrule
        Scholar \cite{iyer2017learningneuralsemanticparser}
          
           & 600 & - & 216 & Single-Turn & Single-Domain & English & 2017 & https://metatext.io/datasets/scholar \\
        \midrule

        Advising \cite{finegan-dollak-etal-2018-improving}
          
           & 4791 & - & - & Single-Turn & Single-Domain & English & 2018 & https://github.com/jkkummerfeld/text2sql-data \\
        \midrule
        EHRSQL \cite{lee2023ehrsqlpracticaltexttosqlbenchmark}
          
           & 5124 & 1163 & 1167  & Multi-Turn & Single-Domain & English & 2023 &  https://github.com/glee4810/EHRSQL\\
        
        \midrule  
         WikiSQL \cite{zhong2017seq2sql}
  
           & 56355 & 8421 & 15878 & Single-Turn & Cross-Domain & English & 2017 & \url{https://github.com/salesforce/WikiSQL?tab=readme-ov-file} \\
        \midrule
        
        Spider                              \cite{yu-etal-2018-spider}  
             
             &  &  &  & Single-Turn  & Cross-Domain & English & 2018 & \url{https://github.com/taoyds/spider} \\

        Spider-SYN \cite{gan2021robustness}  
            
             &   &  &  & Single-Turn  & Cross-Domain & English & 2021 & \url{https://github.com/ygan/Spider-Syn} \\

        Spider-DK \cite{gan2021exploring}  
      
             & &  &  & Single-Turn  & Cross-Domain & English & 2021 & \url{https://github.com/ygan/Spider-DK} \\

        Spider-SS\&CG \cite{gan2022measuring}  
            
             & 7000 & 1034 & 2147  & Single-Turn  & Cross-Domain & English & 2022  & \url{https://github.com/ygan/SpiderSS-SpiderCG?tab=readme-ov-file} \\
        
        Spider-GEN \cite{PatilPKKV23}  
            
             &  &  &  & Single-Turn  & Cross-Domain & English & 2023  & \url{https://github.com/ManasiPat/Spider-Gen} \\
        Spider-Realistic \cite{Zhang2024Benchmarking}  
          
             &  &  &  & Single-Turn  & Cross-Domain & English & 2024  & \url{https://zenodo.org/records/5205322#.YTts_o5Kgab} \\
        Spider-SSP \cite{shaw-etal-2021-compositional}  
          
             &  &  &  & Single-Turn  & Cross-Domain & English & 2021  & \url{https://github.com/google-research/language/tree/master/language/nqg} \\
           
        \midrule      
        KaggleDBQA \cite{lee-etal-2021-kaggledbqa}

             & 272 &-  &-  & Single-Turn  & Cross-Domain & English & 2021  & \footnotesize \url{https://www.microsoft.com/en-us/research/publication/kaggledbqa-realistic-evaluation-of-text-to-sql-parsers} \\
        \midrule   
        BIRD \cite{Wretblad2024Bridging}

             & 8659 &1034  &2147  & Single-Turn  & Cross-Domain & English & 2023  & \url{https://github.com/MohammadrezaPourreza/Few-shot-NL2SQL-with-prompting} \\
        \midrule   
        BigTable-0.2k \cite{Zhang2024Benchmarking}  
       
             & 200 & - & - & Single-Turn  & Cross-Domain & English & 2024  &  - \\
        \midrule  
        DuSQL \cite{wang-etal-2020-dusql}
         
            & 18602 & 2039 & 3156 & Single-Turn & Cross-Domain & English & 2020 & \url{https://paperswithcode.com/paper/dusql-a-large-scale-and-pragmatic-chinese}  \\
        \midrule
        BEAVER \cite{chen2024beaverenterprisebenchmarktexttosql}
         
            & 93 & - & - & Multi-Turn & Cross-Domain & English & 2024 & \url{https://peterbaile.github.io/beaver/}  \\
        \midrule
        CoSQL \cite{yu-etal-2019-cosql}
         
            & 2164 & 292 & 551 & Multi-Turn & Cross-Domain & English & 2019 & \url{https://yale-lily.github.io/cosql}  \\
        \midrule
        ADVETA \cite{pi-etal-2022-towards}
         
            & - & - & - & Multi-Turn & Cross-Domain & English & 2022 & \url{https://github.com/microsoft/ContextualSP}  \\
        \midrule
        CSpider\cite{min-etal-2019-pilot}
         
            & 6831 & 954 & 1906 & Single-Turn & Cross-Domain & Chinese & 2019 & \url{https://github.com/taolusi/chisp}  \\
        \midrule

        TrustSQL \cite{Lee2024TrustSQL}
          
           & - &  -&  -& Single-Turn & Cross-Domain & English & 2024 & https://github.com/glee4810/TrustSQL \\
        \midrule

        SParC \cite{Yu&al.19}
        
           & 9025  & 1203  & 2498 & Multi-Turn & Cross-Domain & English & 2018 & \url{https://github.com/taoyds/sparc} \\

        \midrule

        CHASE \cite{guo-etal-2021-chase}
         
           & 3949  & 755  & 755 & Multi-Turn & Cross-Domain & Chinese & 2021 & \url{https://github.com/xjtu-intsoft/chase} \\
        \midrule

        SQUALL \cite{Shi:Zhao:Boyd-Graber:Daume-III:Lee-2020}
          
           & 9030  & 2246  & 4344 & Single-Turn & Cross-Domain & English & 2020        
           & \url{https://github.com/tzshi/squall} \\
        
	\bottomrule 
	\end{tabular}
	}
	% \vspace{-0.3em}
	\label{dataset}
% \vspace{-1.0em}
\end{table*}

\subsubsection{Single-Domain Datasets}

\textbf{ATIS}\cite{price-1990-evaluation} is derived from the Airline Ticket Subscription System (ATIS), which generates SQL statements from user questions and is a single domain, context-sensitive dataset. 

\textbf{GeoQuery}\cite{10.5555/1864519.1864543} is derived from U.S. Geography, and consists of 880 questions and SQL statements, and is a single domain, context-independent dataset.

\textbf{Scholar}\cite{iyer2017learningneuralsemanticparser} provides a benchmark that reflects the query requirements of real academic databases. The dataset contains 816 annotated natural language queries and corresponding SQL queries, covering a wide range of information retrieval needs in the academic domain. The queries in the dataset cover information about academic papers, authors, citations, journals, keywords, and databases used.

\textbf{Advising}\cite{finegan-dollak-etal-2018-improving} dataset is a Text-to-SQL task assessment dataset focused on student academic advising contexts, drawn from the University of Michigan's course database. Questions are written by students to simulate real questions they might ask during academic advising, and each question is manually annotated with the corresponding SQL query and reviewed by multiple annotators to ensure accuracy and helpfulness.

TPC-DS\cite{10.1145/369275.369291} is a commonly used benchmark in the field of database systems, which, compared to Bird and spider, has a significantly more complex structure of its dataset and is able to model problems in displays more effectively. Even the current state-of-the-art generative AI models fall short in their performance in generating accurate queries on it. \cite{ma2024evaluating}

\subsubsection{Cross-Domain Datasets}

\textbf{WikiSQL}
The two datasets, ATIS \& GeoQuery, have problems such as small data size (less than a thousand SQL sentences) and simple annotation. So, in 2017, Victor Zhong and other researchers annotated 80,654 training data based on Wikipedia, covering 26,521 databases named WikiSQL\cite{zhong2018seqsql}. It posed new challenges to the design of the model, requiring the model to better construct the mapping relationship, make better use of the attributes in the tables, and pay more attention to the decoding process. 

\textbf{Spider}
However, WikiSQL also has a problem; it only involves one table per question and only supports simple SQL operations, which is not very suitable for our daily life scenarios. So, in 2018, researchers at Yale University introduced the Spider\cite{yu-etal-2018-spider} dataset, which is currently the most complex Text-to-SQL dataset. It has the following characteristics:1) The domain is richer, with more than 200 databases from 138 domains. Each database corresponds to 5.1 tables on average, and the databases appearing in the training set and the test set do not overlap.2) The SQL statements are more complex, containing orderBy, union, except, groupBy, intersect, limit, having keywords, and nested queries. The authors divided the SQL statements into 4 levels of difficulty based on their complexity (number of keywords, degree of nesting), and WikiSQL only has EASY difficulty under this division. 

\textbf{KaggleDBQA}
KaggleDBQA\cite{lee-etal-2021-kaggledbqa} is a cross-domain evaluation dataset of real Web databases with domain-specific data types, original formats, and unrestricted questions. It includes 272 examples across 8 databases with an average of 2.25 tables per database. The dataset is known for its real-world data sources, natural problem-creation environment, and database documentation with rich domain knowledge. 

\textbf{DuSQL}
DuSQL\cite{wang-etal-2020-dusql} is a large-scale Chinese dataset designed specifically for cross-domain text-to-SQL tasks, filling the gap of lack of labeled data in the Chinese domain. The dataset manually analyzes real-world problems in several representative applications and contains a large number of SQL queries involving row or column computations.

\textbf{BIRD}\cite{Wretblad2024Bridging} dataset focuses on aspects like grammatical formulation, ambiguity, specificity, and alignment with the database schema. This benchmark aims to bridge the gap between academic research and real-world applications by focusing on the comprehension of database values and the efficiency of SQL queries in large databases. It introduces challenges such as dealing with dirty database contents, requiring external knowledge to link natural language questions with database contents, and ensuring the efficiency of SQL queries. The dataset includes questions of varying difficulty levels—simple, moderate, and challenging. Each question in the dataset is annotated with an optional evidence value, providing context that helps in understanding the query.

\textbf{BEAVER}\cite{chen2024beaverenterprisebenchmarktexttosql} is used to evaluate the performance of large language models in complex SQL generation tasks. Existing publicly available Text-to-SQL datasets (e.g., Spider and Bird) fall far short of real enterprise environments in terms of database structure and query complexity, resulting in large language models that perform well in these tasks but poorly in real-world enterprise environments. The BEAVER benchmark constructs a more representative dataset by anonymizing the data warehouses of the two enterprises that contain complex table joins and aggregation.

\textbf{CoSQL}\cite{yu-etal-2019-cosql} dataset is a cross-domain conversational Text-to-SQL dataset designed for building general-purpose database query dialogue systems. The dataset contains more than 3000 dialogues, more than 30,000 dialogue rounds, and more than 10,000 annotated SQL queries covering 200 complex databases in 138 different domains.CoSQL collects dialogues by means of Wizard-of-Oz (WOZ), where a simulated user on one side and a SQL expert on the other side ask database query questions, and the user builds corresponding SQL queries and returns results.

\textbf{CHASE}\cite{guo-etal-2021-chase} dataset is a large-scale and practical Chinese dataset with cross-database context dependencies. The dataset aims to bridge the gap between existing datasets in terms of context dependency and SQL query complexity. CHASE contains 5,459 problem sequences with a total of 17,940 problems annotated with SQL queries distributed across 280 multi-table relational databases.

\textbf{EHRSQL}\cite{lee2023ehrsqlpracticaltexttosqlbenchmark} is a Text-to-SQL benchmark dataset for Electronic Health Record (EHR) data, aiming at evaluating the real-world applicability of the model in the healthcare domain. The dataset consists of real questions from 222 hospital staff (including doctors, nurses, insurance auditors, etc.), covering common retrieval needs in healthcare scenarios, such as patient information querying, complex statistical computations, etc.
\subsubsection{Augmented-Datasets}

\textbf{ADVETA}\cite{pi-etal-2022-towards} is the first benchmarking dataset specifically designed to evaluate the robustness of Text-to-SQL models under table perturbation. While previous research has focused on perturbation on the natural language side of the problem, ignoring the diversity of tables themselves.

\textbf{Spider-DK}\cite{gan2021exploringunderexploredlimitationscrossdomain} focuses on the ability of the model to work with data obtained from domain-specific knowledge. It transforms the challenge to the use case using data such as Implicit Query Column, Simple Reasoning, Synonym Replacement, and Conditional Generation, etc. Spider-DK is a way to determine if a model has a basic understanding of data and uses old information to process new content. 

\textbf{Spider-SS\&CG}\cite{gan2022measuringimprovingcompositionalgeneralization} aims at realizing tasks through Schema Simplification and Complexity Generation in real databases. As the training progresses, its database is simplified and complicated more. Spider-SS \& CG combines these two elements to check, respectively, the performance of a simplified and complexified database structure.

\textbf{Spider-SYN}\cite{gan2021robustnesstexttosqlmodelssynonym}
 (Synonym Substitution) suggests the introduction of synonym changing to model the synonym variation in the real language. The model will be tested for robustness using a given dataset which will replace schema-related words such as table names and column names with their synonyms. The model's schema linking ability is a problem in this case, as the naming is the opposite of what it should be, e.g., cross-domain tests. 
 
\textbf{Spider-SSP}\cite{shaw-etal-2021-compositional} (Schema-Specific Parsing) concentrates on the parsing oriented towards a schema to test whether the schema generalization is feasible by changing the names of columns and tables in the schema. It insists on schema-dependent parsing capabilities that will be tested against unknown database structures or schema name changes. 

\textbf{Spider-Realistic}\cite{Deng_2021} generates questions and corresponding SQL statements (pairs) that carriers more of a relevant value for real-world applications. The data set is aimed at enterprise performance in real-life databases, which requires the model to work under real-world conditions, particularly in dealing with complex queries that have multiple levels. 

\textbf{CSpider}\cite{min-etal-2019-pilot} addresses the status quo of Chinese as a low-resource language in this task area. Chinese text needs to be processed by word splitting, while SQL keywords and column names of database tables are usually written in English. Word-based semantic parsers are susceptible to word-splitting errors, and cross-language word embedding is very helpful for text-to-SQL mapping.

\textbf{TrustSQL}\cite{Lee2024TrustSQL} aims to evaluate models on their ability to either generate a correct SQL query or abstain from it when the question is unanswerable, making a prediction, or the generated SQL is likely to be incorrect. The data is split in two ways: question-based, which assesses the model's ability to handle different phrasings, and query-based.

 \textbf{BigTable-0.2k}\cite{Zhang2024Benchmarking} built upon the BIRD dataset, which includes various question complexities and dataset sizes. To comprehensively evaluate the performance of LLMs, the authors design five distinct tasks, Text-to-SQL, SQL Debugging, SQL Optimization, Schema Linking, and SQL-to-Text. 

 \textbf{SParC}
 SParC\cite{Yu&al.19} demonstrates complex contextual dependencies, and greater semantic diversity, and requires the model to be able to generalize over unseen domains due to its cross-domain nature and unseen databases used in testing

\section{Methodology}
In this section, we describe in detail the traditional Text-to-SQL methods and their evolution. Traditional methods mainly rely on bi-structured models. These approaches typically use LSTM-based and Transformer-based models to generate SQL queries by learning a contextual representation between natural language questions and database tables. In this context, this paper also discusses a variety of existing frameworks and techniques, including models based on techniques such as graph neural networks, table semantic understanding, and schema linking. These techniques and frameworks further enhance the accuracy and efficiency of Text-to-SQL by improving schema linking, reducing error propagation, and optimizing the use of pre-trained models. 

In addition to the traditional LSTM and Transformer models, the
Pre-trained Models such as BERT, GPT, and T5 have dramatically changed the direction of the Text-to-SQL field. Trained on large-scale text data, these models are able to capture rich semantic representations between natural language and SQL, not only by fine-tuning them to specific tasks but also by combining them with Prompt Engineering to realize applications that do not require large amounts of labeled data. They show high scalability and flexibility in handling complex queries and cross-domain tasks and have become the core of modern text-to-SQL systems. Meanwhile, Fine-tuning optimizes the performance of pre-trained models in specific domains by further supervised training on specific datasets.

Figure \ref{fig:overview} provides a detailed taxonomy of various Text-to-SQL methods, comparing key attributes such as backbone models, optimization strategies, query generation strategies, and datasets used. This table offers a comprehensive overview of different approaches and highlights the role of pre-trained models in modern text-to-SQL systems.

In addition, the introduction of LLM Agents represents a new direction in the development of Text-to-SQL. These systems are not only capable of generating SQL queries, but also interacting with users through multiple rounds of dialog to dynamically adjust the generation strategy. It points out the direction for the further development of Text-to-SQL technology, showing higher flexibility and adaptability. In what follows, we discuss in detail the specific implementation of each approach.

\subsection{Traditional Text-to-SQL Methods}

This part focuses on text-to-SQL models based on traditional deep neural networks. The focus of the discussion is on the model architecture and its application to specific tasks. For instance, the model needs to understand the natural language input and show how it corresponds to the database structure of tables and columns\cite{kumar2022deeplearningdrivennatural, 10.1145/3448016.3457543}.

In the earliest Text-to-SQL tasks, SQL queries were typically generated through a template- or rule-based approach. The model would map the natural language input to a set of predefined SQL templates. Performance is significantly limited when faced with complex database structures and queries. As deep learning techniques matured, more sophisticated methods such as LSTM-based and Transformer-based models began to emerge, offering improved capabilities in handling more complex query scenarios.

The deep neural network system must identify and map out the target, which is the matching data according to the user input. This model generally represents a dual structure, the encoder and the decoder, each of which performs its own duty. The first one, the encoder, is responsible for capturing the semantics of a natural language (NL) question; the second one, the decoder, generates an SQL query based on the representation extracted from the encoder input data\cite{650093}.

\paragraph{LSTM-based Methods}
Traditional methods using LSTM and Transformers generate SQL queries by learning the contextual representation of input natural language (NL) questions and database tables as well.  LSTM-based models were among the first deep learning approaches to be applied to text-to-SQL tasks, as they could effectively model the sequential dependencies between natural language questions and SQL queries. Models such as use Bi-LSTM to learn the semantic representation of question-SQL pairs. While pre-trained models generally outperform these approaches, LSTM-based systems are still employed in some cases. These methods, especially LSTM and its variants, leveraged sequential processing to capture context; however, they faced challenges when dealing with long-range dependencies in complex queries. Traditional techniques work with LSTM \cite{10.1162/neco.1997.9.8.1735} and Transformers \cite{NIPS2017_3f5ee243,raffel2023exploringlimitstransferlearning} by learning contextual representation from input natural language questions and database tables. Models like TypeSQL\cite{yu2018typesqlknowledgebasedtypeawareneural}, Seq2SQL\cite{zhong2017seq2sqlgeneratingstructuredqueries}, SQLNet\cite{xu2017sqlnetgeneratingstructuredqueries}, and SyntaxSQLNet\cite{yu2018syntaxsqlnetsyntaxtreenetworks}  use Bi-LSTM to learn the semantic representation of question-SQL pairs. Pretrained models yield better accuracy compared to these methods. LSTM-based systems, however, are still employed in certain case studies. For example, IRNet\cite{guo2019complextexttosqlcrossdomaindatabase} and RAT-SQL\cite{wang2021ratsqlrelationawareschemaencoding} scheme are introduced that takes advantage of LSTM in their grammar decoder to yield abstract syntax trees.

Although these LSTM-based methods helped in laying the foundation for text-to-SQL research, the limitations in scalability and the ability to generalize across different domains led researchers to explore more advanced architectures, such as Transformer models.

\paragraph{Transformer-based Methods}
A short while ago, works focused on Transformer-based models, complementing them with improved Text-to-SQL performance-oriented structures. These models introduced a new paradigm by employing self-attention mechanisms, which allowed them to handle long-range dependencies more effectively than LSTM models. This shift in architecture significantly improved the models' ability to generate accurate SQL queries, even for complex database schemas. Such settings can be observed. For instance, 
GraPPa\cite{yu2021grappagrammaraugmentedpretrainingtable} introduces grammar-augmented pretraining, which enriches schema understanding.
GAP\cite{shi2020learningcontextualrepresentationssemantic} focuses on contextual representations for tasks of semantic parsing.
StruG\cite{Deng_2021}  presented a new approach called structure objective for text-based table encoding, which emphasizes on the structure of tables.
SCoRe\cite{yu2021score} proposed schema-compound embedding for structured data tasks.
TaBERT\cite{yin2020tabertpretrainingjointunderstanding} was trained to jointly understand the semantics of tabular and textual data using internal pretraining, which resulted in a greater grasp of semantic parsing.
TAPAS\cite{Herzig_2020} pretrains table-based fresh data for tabel data question answering task.
MATE\cite{eisenschlos2021matemultiviewattentiontable} entertains multi-view attention for table learning tasks. 
TableFormer\cite{yang2022tableformerrobusttransformermodeling} suggests a stable transform model for data table comprehension. 
TAPEX\cite{liu2022tapextablepretraininglearning} utilizes table pretraining in terms of logic procedures.
S$^2$SQL\cite{hui2022s2sqlinjectingsyntaxquestionschema} injects syntax into question-schema interactions to improve SQL code generation.
IST-SQL\cite{wang2020trackinginteractionstatesmultiturn} project interaction state in multi-turn SQL tasks. 
At this point, IGSQL\cite{cai-wan-2020-igsql} aids schema linking in intricate SQL queries.
GAZP\cite{zhong-etal-2020-grounded} is a neural approach to a changed zero-shot parsing designed for table question answering.
EditSQL\cite{zhang2019editingbasedsqlquerygeneration}  assembles an editor-like generation method for SQL query formulation. These Transformer-based models introduced innovative techniques such as attention mechanisms and pretraining, which greatly enhanced their performance, particularly in schema linking and complex query generation tasks. These difficulties illustrate various endeavors to improve text-to-SQL performance, where the major three areas are schema linking, propagation of errors, and pre-trained language models-based query generation.

TRANX\cite{Nan2023EnhancingFT} is an artificial intelligence system designed to decipher the syntax and the semantics of the loaded text data. Within the foundation of TRANX, there is a neural abstract syntax parser that is fundamentally the transfer of the natural language into the internal representation. Such as it is in the case here, this is the deep learning practice, where the model is working first by dividing the text into smaller portions and then figuring out the relationships between the individual components. It exploits in-context learning as well by feeding the model its own examples that are in fact frequent in the context it has been introduced to. Notably, the system of architecture has multiple prompt design strategies for improving its performance, one of these strategies is by selecting the examples, which are the most relevant to the input query, and adding extra information on the schema during the task execution.
A number of frameworks and techniques have been outlined in the present day which serve the purpose of improving this sphere. Bridge\cite{lin2020bridgingtextualtabulardata} links and integrates textual and tabular data to provide schema linking with greater efficiency.
SDSQL\cite{hui2021improvingtexttosqlschemadependency}improves schema dependency in Text-to-SQL tasks by leveraging schema dependency structures. 
SLSQL\cite{lei-etal-2020-examining} discusses different techniques used for schema linking in Text-to-SQL models. 
IESQL\cite{ma-etal-2020-mention}  increases the efficiency of mention extraction and schema grounding in Text-to-SQL tasks. 
SEAD\cite{xu2023seadendtoendtexttosqlgeneration} is an end-to-end Text-to-SQL generation method that employs more sophisticated schema linking and is the end-to-end method. 
SmBoP\cite{rubin-berant-2021-smbop} proposes a bottom-up, non-autoregressive method that tackles Text-to-SQL parsing through sequence-dependent probabilistic grammars
DT-Fixup\cite{xu2021optimizingdeepertransformerssmall} fine-tunes Transformer small-based models to enhance their capability for small datasets in Text-to-SQL tasks. 
Relation parsing in SQL is achieved through semi-autoregressive parsing with hybrid relation-aware RasaP\cite{huang2021relationawaresemiautoregressivesemantic} models. 
GNN\cite{bogin2019representingschemastructuregraph} utilizes graph neural networks being employed for the purpose of schema representation for the purpose of improving schema linking. 
ShadowGNN\cite{chen2021shadowgnngraphprojectionneural}  suggests using a graph projection neural network, which is a promising idea as it can boost schema referencing and text-to-SQL accuracy.
SADGA\cite{cai2022sadgastructureawaredualgraph} employs a structure-aware dual-graph structure that provides better schema alignment for Text-to-SQL tasks.
LGESQL\cite{cao-etal-2021-lgesql} uses graph representation learning for schema linking in Text-to-SQL query generation. 
UnifiedSKG\cite{xie2022unifiedskgunifyingmultitaskingstructured} makes it possible to unify multitasking across structured knowledge grounding tasks including Text2SQL. 

These various frameworks and techniques represent significant advancements in the Text-to-SQL domain, particularly in addressing the challenges of schema linking, query parsing, and improving overall model efficiency. By incorporating methods such as schema dependency structures, graph neural networks, and semi-autoregressive parsing, modern Text-to-SQL models are better equipped to handle complex databases and intricate SQL queries. As Text-to-SQL models continue to evolve, the combination of robust schema linking mechanisms, powerful pre-trained models, and advanced parsing techniques promises to bring about more accurate, scalable, and efficient systems capable of handling real-world tasks.

%add reference
\begin{figure*}[htbp]
    \centering
    \vspace{-10pt}
    \includegraphics[height=0.5\textheight]{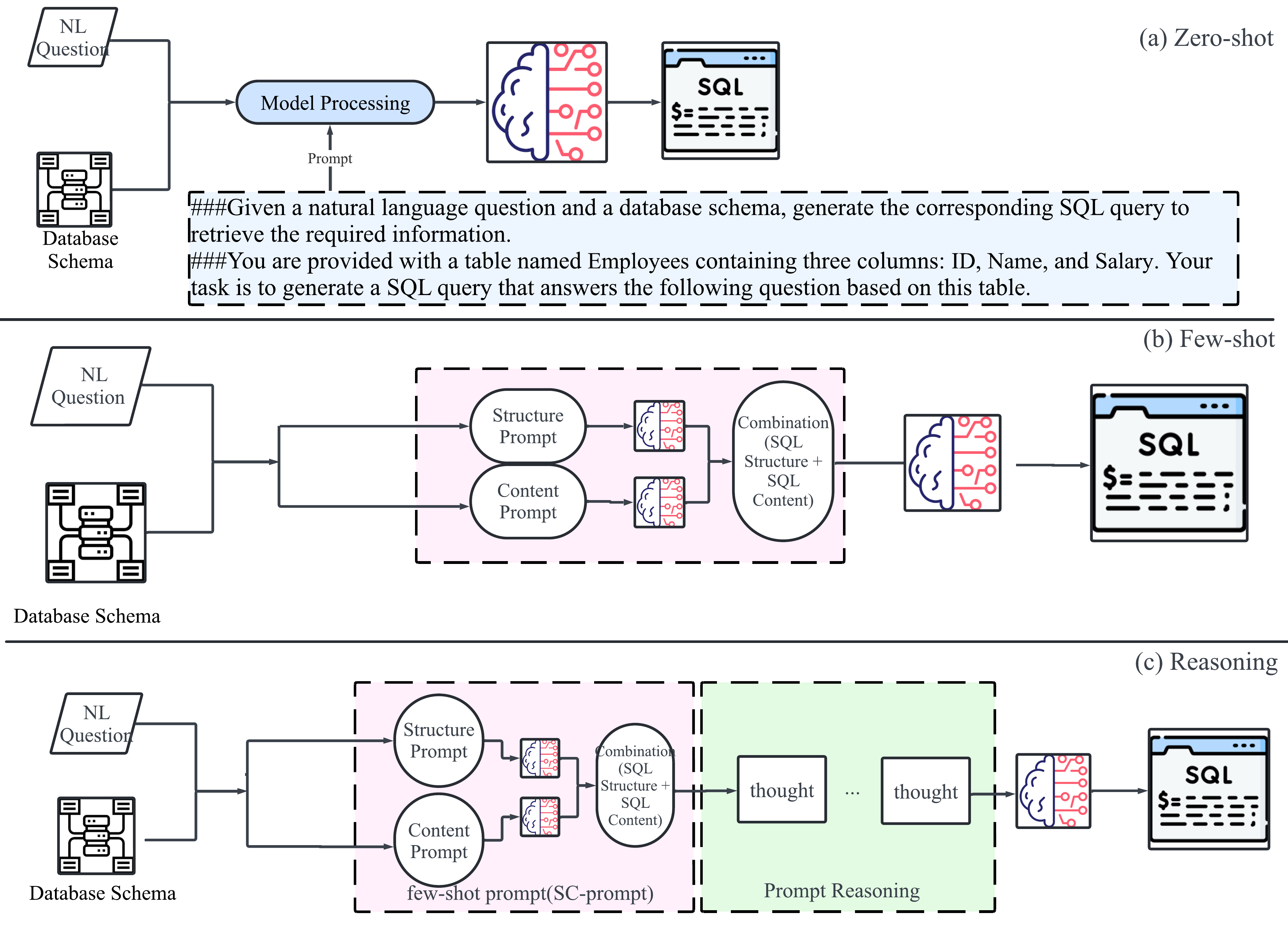} 
    \vspace{-10pt}
    \caption{Prompt Engineering Methods. The figure illustrates three key prompt engineering approaches for Text-to-SQL: (a) zero-shot, where the model generates SQL without prior examples; (b) few-shot, which provides a few examples to guide query generation; (c) Reasoning, breaking down the reasoning process step-by-step for complex queries.} 
    \label{fig:pdfpage}  
\end{figure*}

\subsection{Text-to-SQL with Prompt}

In the process of using prompt engineering, a technique is used to improve the performance of the language models and reinforce the reliability of these models by creating input prompts in detail.  While traditional methods such as LSTM-based and Transformer-based models focus on learning contextual representations through supervised training, prompt engineering offers an alternative approach that leverages the capabilities of large pre-trained language models (LLMs) without requiring additional fine-tuning. This particular method uses the identification of specific prompts or sentences to help guide the model in creating meaningful outputs that are relevant to the intention of the user (See Fig. \ref{fig:pdfpage}). 
Unlike earlier methods that required extensive training on labeled datasets to improve accuracy and generalization, most of the time, Prompt Engineering does not require any training on the model, so it works on the basis of the pre-trained model, which creates SQL directly. The merit of such a technique is in executing results in a short time while not incurring any additional computational charges for model tuning.

This method is vital for problem-solving techniques like hallucinations. The authority and persuasion of the model's answers are heightened by this instruction, since it is grounding the model.  The basic principle of operation is to generate, step by step, the next output token with the highest probability based on the relevant prompts of the input. Therefore, as a contrast to the traditional training-dependent approaches discussed earlier, the core of tackling the Text-to-SQL task with LLMs \cite{huang2024empiricalstudyllama3quantization, huang2024largelanguagemodelinteraction}lies in finding the optimal prompt.
\subsubsection{Zero-shot Prompt}
In the zero-shot approach, the model has zero specific training data about the task. Problems posed directly to the model usually consist of a task description, a test problem, and a corresponding database without any cases. \cite{Rajkumar2022EvaluatingTT,liu2023comprehensiveevaluationchatgptszeroshot, Zhang2024Benchmarking,xue2024dbgptempoweringdatabaseinteractions,xu2024symbolllmfoundationalsymbolcentricinterface} hints at the effect of structure on the zero-shot performance of large language models (LLMs). The model can answer questions accordingly and can make initial judgments based on the large amount of data.
But this method needs both large-scale pre-training language models and a considerable amount of data for adaptation to ensure top-notch quality. In regard to the broader fields of database contents, accuracy can be compromised, and the SQL output can be insufficiently accurate. Nevertheless, this way can be quickly applied to new tasks and areas without spending time on training again. Using this approach method, the output may sometimes be obtained, which could, therefore, be described as having some flexibility variations or unexpected changes.
The techniques of using prompts in the case of zero-second text-to-SQL tasks are dedicated to the work of 
\cite{chang2023promptllmstexttosqlstudy}
\cite{Gao2023TexttoSQLEmpowered}
\cite{zhang2023actsqlincontextlearningtexttosql}
\cite{hu2023chatdbaugmentingllmsdatabases}
\cite{liu2023comprehensiveevaluationchatgptszeroshot}
\cite{liu2023dividepromptchainthought}
\cite{dong2023c3zeroshottexttosqlchatgpt}.

\subsubsection{Few-shot Prompt}

A few-shot is a question posed to the model with only a small number of cases given, and in response, the model generates an answer formulated on the basis of the initiated cases, which results in a better comprehension of the task.
However, here we consider only the cases of good quality. Unlike Zero-shot, Few-shot can significantly yield improved performance of the model, particularly for those tasks characterized by their complexity\cite{li2024flexkbqaflexiblellmpoweredframework}.

\textbf{SC-prompt}\cite{10.1145/3589292 } employs the divide-and-conquer method as a way of tackling the problem of translating the text into two more manageable phases: the structure stage and the content stage. The first stage results in the generation of a basic SQL structure, which contains, but does not limit itself to, the table and column names as their placeholders. It should initially be given a structure where the specific details are expected. The next step consists of replacing the tokens of general description with concrete values. For this purpose, it applies the hybrid approach of text pre-processing that links the static word embeddings to the learnable vectors. 

\textbf{MCS-SQL}\cite{Lee2024MCSSQL} is the method proposed to enhance the precision. Its operation relies on three main activities: schematic linking, parallel generation of SQL, and picking one that meets the purpose. The first stage is schema linking, which later will be performed in two stages: the tables joining and the columns joining. The interaction is repetitive, as the sentence is exchanged in both phases. Afterwards, it links with Schema and generates more than one candidate SQL query. It is through multiple prompts, which aid the exploration of a much wider parameter space, that one is able to accomplish this task. It is the LLM (Language Model) that produces different SQL queries as output, based on those prompts. The primary goal now is to pick the most precise SQL query that matches the input. LLM takes into its consideration the reasoning steps and the scores to choose as the best answer.

 A range of sampling techniques investigating the inspiration demos\cite{nan2023enhancingfewshottexttosqlcapabilities} has been carried out in a selection of declarative programming tasks. It seeks to enhance performance by ensuring that there is an equilibrium between likeness and diversity between demos. The random sampling is then used as a standard to check the effectiveness of the given strategies.

\textbf{SQL-PaLM} \cite{sun2024sqlpalmimprovedlargelanguage} by the first selection strategy of few-shot examples talks about a task of selection both similarity and diversity between various examples. Along with various sampling methods and their combinations, a random sample is taken as a standard for performance evaluation.

\subsubsection{Chain of Thought (CoT)}
The Chain of Thought (CoT)\cite{10.5555/3600270.3602070} prompts activate complex thinking skills with the help of these intermediate steps that are based on reasoning. Its effectivity can be stretched by mixing it with sample-less prompts for complex problem-solving, which implies synthesis before diagnostics. The two key improvements include the enhancement of the reasoning ability in terms of how to notice is the Chain of Thoughts or the CoT principle of the GPT model:
 \cite{Gao2023TexttoSQLEmpowered
, Zhang2024Structure
,hu2023chatdbaugmentingllmsdatabases
,zhang2023actsqlincontextlearningtexttosql
,wang2024dbcopilotscalingnaturallanguage}. 
The Building of a GPT Model was to provide incorrect answers to two claims and give an explanation for the computational approach. In addition, it was found that the model's reasoning performance could be significantly improved by adding key sentences to the prompts, such as "Let's think step by step". The experiments show that such a method helps the rationalization of the GPT even when the relevant samples are not provided. Among these methods are those that provide deep and true insights into the model's reasoning in the case of Text-to-SQL tasks.

Chat2Query\cite{10597681} deploys its zero-shot SQL generation capability, which allows users to input natural language queries and receive SQL outputs without the need for prior model training or domain-specific fine-tuning. The system integrates a text-to-SQL generator, SQL rewriter, SQL formatter, and data-to-chart generator, streamlining the process from query to data visualization. Utilizing the Chain-of-Thought prompt enables step-by-step SQL generation, which improves the accuracy of generated queries, especially in complex or ambiguous situations. The system is built on TiDB Serverless, ensuring scalability and adaptability to different data workloads.

\textbf{ACT-SQL} \cite{zhang2023actsqlincontextlearningtexttosql} delves into the ability of LLM to solve some few-shot learning tasks by examining how the few-shot learning strategy affects the model performance. A hybrid method, based on both the static and the dynamic examples, has been proposed, for example, a selection that is current to test the sample given, and the experimental results showed the efficiency of the chosen strategy.

\subsection{Text-to-SQL with Fine-Tuning}
Fine-tuning is still an important method within LLMs and still offers a high level of improvement over the use of low-cost prompt methods. Fine-tuning methods rely on models that are already pre-trained but are further fine-tuned to fit specific tasks and domains. Depending on the scope of the fine-tuning, it can be categorized in two ways.

\subsubsection{Full-Parameter Fine-Tuning}
Full fine-tuning means fine-tuning all the parameters of the model. In this approach, the entire model is trained with domain-specific data to optimize its performance in a specific task. Full fine-tuning is usually applied in those cases where high accuracy or specific tasks are required. For example, the Text-to-SQL task on the Spider dataset requires the model to generate extremely accurate SQL, in which case all model parameters need to be fine-tuned to improve accuracy.

 \textbf{Knowledge-to-SQL}\cite{Hong2024KnowledgeToSQL}  intends to improve the DELL model's capability to produce relevant knowledge more swiftly, thus giving an added boost to the performance of these Text-to-SQL systems. Semantic techniques are adopted to find the best matching table to the query. First, supervised fine-tuning is performed, and then the model is refined by applying the Direct Preference Optimization (DPO) algorithm.

\textbf{SGU-SQL}\cite{Zhang2024Structure} is a system depending on the structure of the questions and schema. To begin with, user queries and databases are linked through an enriching framework. This connection can be made through graph-based structures, and the system SGU-SQL will decompose the complicated, interrelated structures through grammar trees. The system is also based on a specially designed structure-based linking mechanism that connects the query structure at the node level. The process begins with the composition of the node representations, and later, it is directed by the propagation of messages, which self-structures and, therefore, makes the model capture the main relationships. Ultimately, the structure-blindness measure is utilized by the model to merge the schema graph and query graph, and afterward, the merged information is transferred to the adjacent graph.

\textbf{DIN-SQL}\cite{Pourreza2023DINSQ} breaks down the complex task of converting test to SQL queries into smaller, manageable sub-tasks, which helps LLMs perform better. Firstly, its schema linking module identifies references to database schema and condition values in NL queries. It will classify each query into one of three classes: easy, non-nested complex, and nested complex. This classification helps in using different prompts for each query class. For complex queries, it includes an intermediate representation called NatSQL, which simplifies the transition from natural language to SQL by removing certain SQL operators that do not have clear counterparts. Then the SQL generation module generates the final results based on the solutions of sub-tasks. After generating the initial query, the self-correction module reviews and corrects any errors. The new method achieved the state-of-the-art on the Spider dataset and BIRD benchmark with the accuracy of 85.3\% and 55.9\%.

\textbf{MAC-SQL}\cite{wang2024macsqlmultiagentcollaborativeframework} could be much more efficient in its work of query generation for SQL by involving multiple intelligences: the fundamental Decomposer intelligence, the Selector intelligence, and the Refiner intelligence. The Decomposer Intelligence, when given a complex SQL query to resolve, breaks it into fragment sub-problems sequentially through chained reasoning and generates the final SQL query in a stepwise manner. The usage of Selector Intelligence limits the database to eliminate the problem-specific data interference, while Fixer Intelligence repairs mistaken SQL via SQL execution using outside tools. The MSCFT is the method fine-tuning strategy used during SQL-Llama development, and this model is an open-source version based on the Code Llama. This model is trained using the dataset of instructions generated from the MAC-SQL intelligence tasks.

Few-shot learning is combined with Instruction Fine-Tuning that further extends the improvement in the large language model's performance in the text-to-SQL tasks. First of all, \cite{sun2024sqlpalmimprovedlargelanguage} proposes consistent decoding and instruction execution-based filtering of errors using a few sample hints. It investigates instruction fine-tuning to widen the coverage of the training data, apply data augmentation, and integrate specific content to the queries. This article also proposed a methodology, although with selection at test time, to further boost the accuracy by combining output and execution feedback in different paradigms.

\textbf{Symbol-LLM} \cite{xu2024symbolllmfoundationalsymbolcentricinterface} employs a two-stage supervised fine-tuning framework for this purpose. This strengthens the symbolic processing capability of large models. In the first phase (injection), the model acquires the inherent dependencies in the symbolic data based on fine-tuning (SFT) of an underlying model. Thus, these latter became the new optimizators (MLE) in order to handle symbolic tasks. During the fusion period, the model is not only enhanced by the symbolic and natural language data but is also boosted with the combined data.

\begin{table*}[ht]
\caption{Taxonomy of text-to-SQL Methods.}
\centering
\large
\renewcommand{\arraystretch}{1.3}
\setlength{\tabcolsep}{12pt} 
\resizebox{\textwidth}{!}{%
\begin{tabular}{|p{2cm}|p{1.4cm}|p{2cm}|>{\centering\arraybackslash}p{0.9cm}|p{4.5cm}|p{4cm}|p{3cm}|p{3cm}|p{2.5cm}|>{\centering\arraybackslash}p{2cm}|}

\hline
\textbf{Methods}      & \textbf{Released Time} & \textbf{Backbone Models}   & \textbf{Access} & \textbf{Optimization Strategy}                            & \textbf{Query Generation Strategy}                   & \textbf{Robustness and Error Handling}          & \textbf{Dataset}                                         & \textbf{Metrics} & \textbf{Schema Linking} \\ \hline
SC-prompt             & Jun-23                 & T5                         & \ding{51}                  & Task Decomposition                                        & Guided Decoding          & -                                                   & Spider, CoSQL, GenQuery                              & EM, EX            & \ding{55}                   \\ \hline
MCS-SQL               & May-24                 & GPT-4                      & \ding{55}                & Prompt Tuning                              & Guided Decoding          & Self-Consistency                                    & Spider, Bird                                       & EX, VES           & \ding{51}                  \\ \hline
SQL-PaLM              & May-23                 & Palm-2                     & \ding{55}                & Prompt Tuning                              & Consistency Decoding                                 & Self-Correction, Self-Debugging                     & Spider, Spider, Bird-SYN, Spider-DK, Spider-realistic & EX, TS            & \ding{51}                  \\ \hline
ACT-SQL               & Oct-23                 & -                          & -                            & Chain of thought (CoT)                                    & Greedy Search                                        & Self-Correction                                    & Spider, SParC, CoSQL                                & EM, EX, TS         & \ding{51}                  \\ \hline
Chat2Query            & May-24                 & -                          & -                            & Chain of thought (CoT)                                    & -                                                   & -                                                   & Spider                                             & EM, EX, VES        & \ding{55}                   \\ \hline
Knowledge-to-SQL      & Feb-24                 & LLaMA-2-13b                & \ding{51}                  & Expert Fine-Tuning            & Framework-Based                        & -                                                   & Spider, Bird                                        & EX, VES           & \ding{55}                   \\ \hline
SGU-SQL               & Feb-24                 & GPT-4                      & \ding{55}                & Expert Fine-Tuning            & Guided Decoding          & -                                                   & Spider, Bird                                        & EX, EM            & \ding{51}                  \\ \hline
DIN-SQL               & Apr-23                 & GPT-4                      & \ding{55}                & Task Decomposition                                        & Greedy Search                                        & Self-Correction                                    & Spider, Bird                                        & EX, EM            & \ding{51}                  \\ \hline
MAC-SQL               & Dec-23                 & GPT-4                      & \ding{55}                & Task Decomposition                                        & Greedy Search                                        & Refiner                                             & BIRD                                 & EX, EM, VES        & \ding{51}                  \\ \hline
Symbol-LLM            & Nov-23                 & LLaMA-2-Chat               & \ding{51}                  & Expert Fine-Tuning                     & Greedy Search                                        & -                                                   & Spider, SParC, CoSQL                                & EM                & \ding{55}                   \\ \hline
DAIL-SQL              & Aug-23                 & GPT-4                      & \ding{55}                & Supervised Fine-tuning                                    & Greedy Search                                        & Self-Consistency                                    & Spider, Spider-Realistic                            & EX, EM            & \ding{55}                   \\ \hline
CLLMs                 & Feb-24                 & Deepseek-coder-7B-instruct & \ding{51}                  & Performance \& Efficiency Enhancements                    & Greedy Search                                        & -                                                   & Spider                                             & EX                & \ding{55}                   \\ \hline
StructLM              & Feb-24                 & CodeLlama-Instruct         & \ding{51}                  & Expert Fine-Tuning            & -                                                   & -                                                   & Bird, CoSQL, SParC                                  & EX, EM            & \ding{51}                  \\ \hline
SQL-GEN               & Aug-24                 & MoE                        & -                            & Expert Fine-Tuning                     & -                                                   & -                                                   & Bird                                              & EX                & \ding{51}                  \\ \hline
CodeS                 & Feb-24                 & StarCoder                  & \ding{51}                  & -                                                         & Beam Search                                          & Execution-Guided SQL Selector                      & Spider, Bird                                       & EX, TS            & \ding{51}                  \\ \hline
RESDSQL               & Feb-23                 & T5                         & \ding{51}                  & Skeleton Parsing                                          & Beam Search                                          & Execution-Guided SQL Selector                      & Spider-DK, Spider-Syn, and SpiderRealistic           & EM, EX            & \ding{51}                  \\ \hline
Tool-SQL              & Aug-24                 & GPT-4                      & \ding{55}                & Query Error Handling                                      & Python Interpreter                                   & -                                                   & Spider, Spider-Realistic                            & EX, EM            & \ding{51}                  \\ \hline
SQLFixAgent           & Jun-24                 & GPT-3.5-turbo              & \ding{55}                & Query Error Handling                                      & Perturbation-Based Query Generation                  & Refiner                                             & Spider, Bird, Spider-SYN, Spider-DK, Spider-realistic & EX, EM, VES        & \ding{51}                  \\ \hline
MAG-SQL               & Aug-24                 & -                          & \ding{55}                & Query Error Handling                                      & -                                                   & Refiner                                             & Spider, Bird                                        & EX, VES           & \ding{51}                  \\ \hline
MAGIC                 & Jun-24                 & GPT-4                      & \ding{55}                & Expert Fine-Tuning                     & -                                                   & Self-Correction                                    & Spider, Bird                                        & EX, VES           & \ding{55}                   \\ \hline
SuperSQL              & Jun-24                 & GPT-4                      & \ding{55}                & -                                                         & Greedy Search                                        & Self-Consistency                                    & Spider, Bird                                        & EX, EM            & \ding{51}                  \\ \hline
\end{tabular}%
}
\end{table*}

\subsubsection{Parameter-Efficient Fine-Tuning}
Parameter-efficient fine-tuning fine-tunes only some parameters of the model, usually some specific layers or modules. Such fine-tuning can effectively reduce training time and computational resource consumption while maintaining high performance. This approach usually targets domain-specific texts or structures, preserving the general linguistic knowledge already learned in the pretrained model and optimizing only for subtle differences in the Text-to-SQL task.

For example, when it is about fine-tuning the complexity of SQL statements or the database schema referred to as schema, only the layers or parameters that pertain to understanding of the schema need to be retrained, hence saving the model's training cost and nothing else. The advantage of this way of doing things is reducing the training overhead on the one hand and achieving a happy result of the compromise between effectiveness and efficiency on the other hand. 
\cite{chen2023acceleratinglargelanguagemodel} explains the proposed idea of accelerating the decoding process of a large language model by suggesting sample sequences. The technique does not modify the current model, but instead develops a rough instrument that produces possible candidates; the decision of rejection sampling ensures that the output distribution of the model stays unchanged.

In the inference process of \cite{leviathan2023fastinferencetransformersspeculative}, some computational steps can be approximated by smaller, more efficient models. Particularly, using a quite involved approximation model, many candidate tokens are generated and based on the trained target model, tokens are verified in parallel to find the ones that are within the target distribution of the model. Once done, the distance from the goal is reduced, leading to quicker model moves that finally end the decoding process.

\cite{song2022learningnoisylabelsdeep} mentions some techniques to enhance the robustness of models in noisy data by fine-tuning the manually selected pre-trained models and applying robust Regularization (RR). The provided regularization methods, such as fine-tuning using pre-trained models, can help to increase modeling robustness effectively. In pre-training of the model, providing it clean data and then restoring it on noisy data, the model is less likely to avoid design overfitting as a result of noisy labels. Another way is to add an explicit regularization to the pre-training technique mentioned here, and further noise robustness improvement will be achieved, for instance, in the case of PHuber, which is a fine-tuning additionally.

\textbf{DAIL-SQL} \cite{Gao2023TexttoSQLEmpowered} presents an exploration of the effect of Supervised Fine-Tuning (SFT). The technique is a systematic approach of SFT by employing LLMs to be fine-tuned for specific Text-to-SQL tasks trained by task-specific training data. It investigates the representation strategies used for supervised fine-tuning with insights from different strategies on the efficiency of supervised fine-tuning.

The paper \cite{deng-etal-2022-recent} has various avenues for operations by employing pre-trained models and customizing the pre-trained ones. As for example, through a method of language pre-training called BERT\cite{devlin-etal-2019-bert}, natural languages and database schemas are encoded with the goal of a deeper representation of syntactic and semantic structures.

Moreover, the pre-trained model embeddings are customized using adaptive fine-tuning to respond to the schema of the database or the respective language problem. For instance, SQLova\cite{hwang2019comprehensiveexplorationwikisqltableaware} and X-SQL\cite{he2019xsqlreinforceschemarepresentation} utilize multiple post-training, self-organized semi-supervised learning to improve performance. Fine-tuning of other models: through pre-training the models on tabular data and further fine-tuning in the Text-to-SQL task, improvements in the accuracy and robustness of models like TaBERT\cite{yin-etal-2020-tabert} and Grappa\cite{yu2021grappa} take place.

\textbf{CLLMs} \cite{kou2024cllmsconsistencylargelanguage} coupling is additionally optimized by updating with the losses of the two streams: consistency loss and autoregressive loss (AR loss). Consistency loss is employed to ensure that the model settles to a final state with minimum energy across any kind of input, increasing the rate of convergence. Through learning the Jacobi trajectories of the model, it is able to generate multiple markers via one iteration step, and thus it needs fewer calculations to output the current results.

\textbf{StructLM} \cite{zhuang2024structlmbuildinggeneralistmodels} optimizes pretrained models through Instruction Fine-Tuning. Instruction Fine-tuning combines structured data with generic instruction-tuning data to enhance the model's generalization ability on structured knowledge tasks. In addition, the paper explores fine-tuning on top of code pre-training and finds that code pre-training has a significant enhancement effect on processing structured knowledge tasks.
\cite{Roberson2024Effectiveness,zhou2024dbgpthubopenbenchmarkingtexttosql} uses LoRA\cite{hu2021loralowrankadaptationlarge} and QLoRA\cite{dettmers2023qloraefficientfinetuningquantized} fine-tuning to reduce memory requirements and adapt to SQL generation tasks. and provides a standardized set of evaluation pipelines.LoRA freezes the weights of the pre-trained model and injects trainable layers in each Transformer block for efficient fine-tuning and reduced memory footprint. 

\cite{jang2023exploringbenefitstrainingexpert} investigates the performance of language models on task generalization. It was found that expert language models trained for a single task can outperform multitask instruction fine-tuning (MT) language models on unseen tasks. The expert language model avoids negative task transfer and catastrophic forgetting and performs well in learning new tasks compared to the MT model. Through independent training and the Retrieval of Experts (RoE) mechanism, the study demonstrates the potential for selecting appropriate expert models in multi-task scenarios. It uses a Parameter-Efficient Fine-Tuning approach, which reduces training costs and improves efficiency by freezing the underlying pretrained language model and fine-tuning only the addition of extra adapters.

\subsection{Text-to-SQL with Task-Training}

Such a strategy involves fine-tuning and training the complete models from scratch for the specifics of the work. Differing from the two mentioned approaches, Fully Pre-Trained Models range from those off-the-shelf pre-trained language models that are now available but are eclectically selected for the SQL generation tasks. Typically, that includes deep learning methods, such as CODES models which are based on CNN architecture, Mixture of Experts (MoE) models, and Transformers-based models\cite{meng2023masseditingmemorytransformer}.

\subsubsection{Mixture of Experts Models}

MoE model is a new type of architecture that works by introducing several expert modules, and, at each module, one is responsible for a specific type of task or input. The MoE model of text-to-SQL tasks will allow multiple experts to be involved in the different modules of tasks, such as natural language understanding, database schema parsing, and SQL generation, thus making it easier for the system to learn.

\textbf{SQL-GEN}\cite{pourreza2024sqlgenbridgingdialectgap} adopts the LLM technique to extend the existing SQL templates and create a tutorial-driven SQL engine that serves purely as a code interpreter for different dialects. The strategy boosts performance on BigQuery and PostgreSQL dialects remarkably. In addition, the paper also recommends an MoE (Mixture of Experts) approach, which is used generally to boost the performance of such models further by combining the DB-specific models into a single system with the help of dialect keyword injections.
\subsubsection{Transformer-based Models}
The Transformer model is a mainstream architecture in the field of deep learning in recent years, which is particularly suitable for generating complex SQL queries by modelling long-distance dependencies through the attention mechanism. Fully trained Transformer models are built from scratch and trained with large amounts of text and SQL data and can well handle cross-domain and complex SQL generation tasks. Such models can show strong generalization capabilities in database query scenarios, especially in cross-domain and multi-language environments.

\textbf{CodeS}\cite{Li2024CodeS} is specifically designed for Text-to-SQL tasks. The paper points out that while existing large-scale language models (e.g., GPT-4, ChatGPT) perform well in the task of generating SQL, their closed-source nature poses data privacy risks and high inference costs. Accordingly, CodeS has become an open-source alternative that seeks to achieve effective and accurate SQL generation by reducing the number of parameters (1B to 15B) and a pre-trained SQL architecture that is fine-tuned on SQL generation tasks. Therefore, it applies database hints that filter relevant tables, columns, and values, such as BM25, to generate accurate SQL queries. Regarding the new domain adaptation, CodeS generates automatically a huge amount of (question, SQL) pairs through the data augmentation techniques, bi-directional. 

\textbf{MIGA}\cite{Fu2023MIGA} is based on PLMs like T5, which is highly proficient in the Text-to-SQL conversion by breaking the main goal into a bunch of smaller but interconnected sub-goals and recognizing the relationships learned during pre-training, so that a Seq2Seq formatted Text-to-SQL could be generated. The three sub-tasks—Related Schema Prediction (RSP), Turn Switch Prediction (TWP), and Final Utterance Prediction (FUP)—are implemented to boost the model accuracy and intelligence by exposing it specifically to the aforementioned SQL generation aspects. Moreover, it is imperative for MIGA to introduce four types of SQL perturbations that aim to minimize dependencies on previously created SQL instructions and to solve the problems of error propagation. Therefore, the model's accuracy will be enhanced, as it will be able to process diverse conversational inputs in a more robust fashion. When plaguing the model with specific languages, the model is guided by these specific language prompts, which are based on the T5 training methodology. This method involves combining a specific task prompt to a training sample pertaining to an input to a given target task.

\textbf{SQLova}\cite{hwang2019comprehensiveexplorationwikisqltableaware} presents significant progress observed in both execution accuracy and logical form accuracy. Taking advantage of BERT for embedding form text and further generation of SQL queries through multi-layer LSTM, the model applies form context as well. It must be apparent that the SQL generation master-theory, which integrates form-aware BERT and execution-guided decoding (EG) methods, achieves excellent results against the SQL.

\textbf{RESDSQL} \cite{Li2023RESDSQL} enhances the Text-to-SQL task by decoupling schema linking from skeleton parsing, in which the first process is to identify relevant database tables and the other is to establish the SQL queries structure. This helps to simplify the process of creating accurate SQL queries. This model features an enhanced encoder that prioritizes the most relevant.

\cite{rai2023improvinggeneralizationlanguagemodelbased} introduces two approaches to improve generalization in Text-to-SQL semantic parsing based on pre-trained language models. Token Preprocessing helps the language model's tokenizer generate semantically meaningful tokens by processing the naming conventions of database schemas and SQL queries. This includes converting serpentine nomenclature to a natural naming format, handling column references for dot notation, and extending the spelling of some SQL keywords. Component Boundary Marking inserts special markers at aligned component boundaries between NL input and SQL output. These markers are inserted in the input and output to help the model identify semantic component boundaries and enhance the language model's ability to generalize to combinations.

\cite{fürst2024evaluatingdatamodelrobustness} explored the performance of different data models under real user queries Two specific approaches were used. valueNet: a BART-based encoder using Intermediate Representation (IR) to transform natural language into SQL. the architecture and mechanisms of IRNet were used. in particular, connecting entities in a natural language problem to tables in the database through Schema linking, columns, and data values. There is also T5-Picard, which is based on a variant of the T5 model that incorporates the Picard method\cite{scholak2021picardparsingincrementallyconstrained}, which is an incremental parsing method that constrains the decoded output of the language model to valid SQL statements.
\cite{parthasarathi2023conversationaltexttosqlodysseystateoftheart} introduces multi-task training, where different Text-to-SQL tasks (CoSQL, Spider, SParC) are combined and discrete task-specific prompts are used Two reordering methods are employed: the Query Plan (QP) Model: generates SQL queries with predictions of whether the query should contain specific SQL clauses. and Schema Linking (SL): improves SQL generation by performing Schema Linking on the dialog context, ensuring that column and table names in SQL queries are consistent with natural language input.

\cite{guo2023promptinggpt35texttosqldesemanticization} enhances prompts primarily through de-semanticization and skeleton retrieval. Galois\cite{saeed2023queryinglargelanguagemodels} extracts structured data from LLMs by decomposing SQL queries into multiple steps and converting these steps into natural language prompts, it enhances the process primarily through de-semanticization and skeleton retrieval. Not only the queries from LLMs but also traditional databases can be processed using this technique that allows users to query the information contained in LLMs with SQL commands.

\subsection{Text-to-SQL with LLM Agent}
Intelligent agent-based Text-to-SQL systems have become a cutting-edge solution for dealing with complex SQL query generation tasks. By collaborating with multiple intelligences, the LLM Agent framework not only automatically generates SQL queries, but also dynamically adapts and corrects SQL statements, handles database matching issues, and improves query accuracy and execution through external tools. By introducing the mechanism of multi-intelligentsia collaboration, these systems are equipped with stronger flexibility and adaptive capability to decompose complex tasks, detect and repair errors, and optimize query conditions. This paper discusses several LLM Agent systems that dramatically improve the performance and reliability of Text-to-SQL tasks through stepwise reasoning, external supervision, and query optimization.

\textbf{MAC-SQL}\cite{wang2024macsqlmultiagentcollaborativeframework} could be much more efficient in its work of query generation for SQL by involving multiple intelligences: the fundamental Decomposer intelligence, the Selector intelligence, and the Refiner intelligence. The Decomposer Intelligence, when given a complex SQL query to resolve, breaks it into fragment sub-problems sequentially through chained reasoning and generates the final SQL query in a stepwise manner. The usage of Selector Intelligence limits the database to eliminate the problem-specific data interference, while Fixer Intelligence repairs mistaken SQL via SQL execution using outside tools. The MSCFT is the method fine-tuning strategy used during SQL-Llama development, and this model is an open-source version based on the Code Llama. This model is trained using the dataset of instructions generated from the MAC-SQL intelligence tasks.

\textbf{Tool-SQL} \cite{wang2024tool} proposes a helper-agent framework that equips LLM-based agents with two specialized tools, a retriever and a detector. They are responsible for diagnosing and correcting SQL queries that suffer from database mismatch problems. The Retriever helps LLM-based agents verify the correctness of SQL conditional clauses by aligning the values in the SQL query to the corresponding cells in the database.

\textbf{SQLFixAgent}\cite{cen2024sqlfixagent} adopts a multi-agent collaborative approach and consists of three main agents, first SQLRefiner is responsible for generating the core agent for the final repaired SQL query. Secondly, SQLReviewer detects syntactic and semantic errors in SQL queries. Generate multiple candidate SQL statements using a fine-tuned SQLTool.

\textbf{MAG-SQL}\cite{xie2024mag} consists of four parts. First, it performs a soft-column selection of the database schema, the purpose of which is to filter out redundant information. Then in the next module, it breaks down the problem into a series of smaller problems, feeds into the Iterative Generation module, which iteratively generates sub-SQL for each problem, and then a refiner executes these queries using external tools to optimize any incorrect SQL queries. External supervision is used throughout the process to ensure that the generated queries are consistent with the expected queries

\textbf{MAGIC}\cite{askari2024magic} proposes a novel agent approach to automatically create self-correcting guides by iterating over incorrect queries in the text-to-SQL process. By providing automated, data-driven to mimic the correction process of human experts.

Distyl AI's Analytics Insight Engine \cite{maamari2024end} generates complex queries in arbitrary environments and is able to improve the generated results through user feedback and dynamics. It also performs external knowledge retrieval, where the user's question is paraphrased into an authoritative form that captures the user's intent and, based on that intent, proceeds to retrieve relevant examples.

\textbf{SuperSQL}\cite{li2024dawn} uses a number of approaches to different architectures. In the preprocessing phase, it uses architectural links in RESDSQL and database content in BRIDGE v2 as a way to enhance connectivity to the overall database, uses DAIL-SQL's few-shot prompt engineering module to select contextual examples based on similarity and self-consistency to ensure the reliability of the generated content, and then generates SQL queries using the SQL query generation using greedy decoding strategy

\section{Conclusion}
% \section{Text}
In this paper, we thoroughly review and analyze the implications of large-scale language models (LLMs) in the text-to-SQL framework, looking for further research from different perspectives. To start with, the contribution of this paper lies in classifying the application of LLMs on the task of text-to-SQL into two groups: traditional approaches and approaches to hint engineering on the one hand, and fine-tuned approaches on the other. We will develop in-depth important aspects, such as the general structure of hints, ways of supplementing knowledge, example choosing, and reasoning.

We are looking forward to future research being able to expand the methods that allow the generation to be smarter and more cost-effective and generation smarter and more cost-effective SQL query generation that shows stronger generalization ability in cross-domain and cross-language application scenarios. Continuous optimization and innovation in the Text-to-SQL field imply that more efficient, high-performance decision-making and data information retrieval procedures via formal language will be realized, marked by further unprecedented developments in the relevant fields.
\newpage
\newpage

\bibliographystyle{IEEEtran}
\bibliography{reference}

\vfill

\end{document}